\documentclass[pra,aps,twocolumn,eqsecnum,superscriptaddress,showpacs,amsmath]{revtex4-1}
\usepackage{amsmath,amssymb,graphicx,graphics,color}
\usepackage{dcolumn}
\usepackage{bm}
\usepackage{psfrag}
\usepackage{graphicx}
\usepackage{latexsym}
\usepackage{epsfig}
\usepackage{color}
\usepackage{bm}
\usepackage{amsmath}
\usepackage{subfigure}

\begin{document}

\title{Model analysis of magnetic susceptibility of Sr$_2$IrO$_4$ - 2D $J_{\rm eff}$ = 1/2 Heisenberg system with competing interlayer couplings}
\author{Tomohiro Takayama}
\affiliation{Max Planck Institute for Solid State Research, Heisenbergstrasse
  1, 70569 Stuttgart, Germany}
\affiliation{Department of Physics and Department of Advanced Materials,
  University of Tokyo, 7-3-1 Hongo, Tokyo 113-0033, Japan}
\author{Akiyo Matsumoto}
\affiliation{Department of Physics and Department of Advanced Materials,
  University of Tokyo, 7-3-1 Hongo, Tokyo 113-0033, Japan}
\author{George Jackeli} 
\affiliation{Max Planck Institute for Solid State Research, Heisenbergstrasse
  1, 70569 Stuttgart, Germany}
\affiliation{Institute for Functional Matter and Quantum Technologies, University of Stuttgart, Pfaffenwaldring 57, 70569 Stuttgart, Germany}
\author{Hidenori Takagi}
\affiliation{Max Planck Institute for Solid State Research, Heisenbergstrasse
  1, 70569 Stuttgart, Germany}
\affiliation{Department of Physics and Department of Advanced Materials,
  University of Tokyo, 7-3-1 Hongo, Tokyo 113-0033, Japan}
\affiliation{Institute for Functional Matter and Quantum Technologies, University of Stuttgart, Pfaffenwaldring 57, 70569 Stuttgart, Germany}
\date{\today}
\begin{abstract}
We report the analysis of magnetic susceptibility $\chi$($T$) of Sr$_2$IrO$_4$ single crystal in the paramagnetic phase. We formulate the theoretical susceptibility based on isotropic Heisenberg antiferromagnetism incorporating the Dzyaloshinsky-Moriya interaction exactly, and include the interlayer couplings in a mean-field approximation. $\chi$($T$) above $T_{\rm N}$ was found to be well described by the model, indicating the predominant Heisenberg exchange consistent with the microscopic theory. The analysis points to a competition of nearest and next-nearest neighbor interlayer couplings, which results in the up-up-down-down configuration of the in-plane canting moments identified by the diffraction experiments. 
\end{abstract}


\maketitle
\section{Introduction}

 Complex iridium oxides recently emerged as a novel playground for correlated electron physics where strong spin-orbit coupling of 5$d$ Ir, comparable to its modest Coulomb $U$, plays a critical role to produce unprecedented electronic phases. A notable example is the spin-orbital Mott state with local $J_{\rm eff}$ = 1/2 wave function produced by the interplay between spin-orbit coupling and Coulomb $U$. The $J_{\rm eff}$ = 1/2 wave function consists of equally weighted superposition of three $t_{2g}$ orbitals with imaginary components, $|J_{\rm eff} = \pm1/2> =  \frac{1}{\sqrt{3}} \{|d_{xy},\pm\sigma> \pm |d_{yz},\mp\sigma> + i|d_{zx},\mp\sigma>\}$ where $\sigma$ denotes the spin state \cite{BJ_PRL2008}.  The $J_{\rm eff}$ = 1/2 Mott state was first identified in the K$_2$NiF$_4$-type layered perovskite Sr$_2$IrO$_4$ \cite{BJ_Science2009}. In spin-orbital Mott insulators, the magnetic coupling between $J_{\rm eff}$ = 1/2 isospins is mediated by their direct overlap or superexchange interaction via anions, and is therefore critically affected by the unique form of $J_{\rm eff}$ = 1/2 wave function.

The magnetic coupling between $J_{\rm eff}$ = 1/2 isospins was studied theoretically in Ref.~\cite{Jackeli_PRL2009}, and the low energy Hamiltonian was constructed. In the case of $90^{\circ}$ bond of Ir-O-Ir, where the IrO$_6$ octahedra share their edges, the destructive interference manifests itself in the two superexchange paths of Ir-O$_2$-Ir plaquette owing to the imaginary components of $J_{\rm eff}$ = 1/2 state. As a consequence, the magnetic exchange takes the form of an anisotropic bond-dependent interaction. Such bond-dependent coupling gives rise to strong frustration when iridium ions are placed on a tri-coordinated motif like honeycomb lattice, invoking a possible route for Kitaev spin liquid \cite{Kitaev_AP}. In contrast, for 180$^{\circ}$ bond of Ir-O-Ir, relevant to Sr$_2$IrO$_4$, the magnetic coupling is proposed to comprise isotropic Heisenberg exchange and pseudodipolar interaction stemming from Hund's coupling. The emergence of isotropic Heisenberg coupling, rooted in the isotropic $J_{\rm eff}$ = 1/2 wave function, is rather unexpected since spin-orbit coupling is generally considered to produce magnetic anisotropy.

The presence of Heisenberg coupling was indeed found experimentally in Sr$_2$IrO$_4$. Sr$_2$IrO$_4$ undergoes a magnetic transition around $T_{\rm N} \sim$ 240 K \cite{Cao_PRB1998}. A resonant x-ray diffuse scattering showed that the two-dimensional (2D) magnetic correlation survives in the IrO$_2$ planes above $T_{\rm N}$ \cite{Fujiyama_diffuse}, and the temperature dependence of correlation length obeys the relation theoretically proposed for the 2D $S$ = 1/2 isotropic Heisenberg antiferromagnetism (IHAF) on a square lattice \cite{CHN1989}. The nearly gapless magnon dispersion observed by resonant inelastic x-ray scattering (RIXS) is consistent with those expected for IHAF \cite{BJ_RIXS2012}. The 2D IHAF of Sr$_2$IrO$_4$ is reminiscent of the isostructural compound La$_2$CuO$_4$, a parent Mott insulator of high-$T_{\rm c}$ superconductor with 2D $S$ = 1/2 IHAF \cite{Aharony review, Keimer_PRB1992}. The similarity of two compounds led the theoretical prediction of possible superconductivity in Sr$_2$IrO$_4$ upon doping \cite{FWang_PRL2011, Yunoki_SC, Kee_SC} and the observation of Fermi arcs and $d$-wave gap on the doped surface of Sr$_2$IrO$_4$ \cite{BJ_Fermi arc, d-wave_1, d-wave_2}.

\begin{figure}
\epsfysize=80mm
\includegraphics[width=8cm, clip]{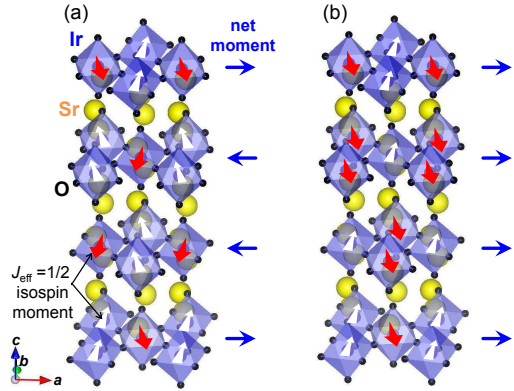}
\caption{Crystal and magnetic structures of Sr$_2$IrO$_4$ (a) at zero magnetic
  field, and (b) above metamagnetic critical field $\mu_{0}H_{\rm c} \sim 0.15$
  T. Yellow, blue and black spheres represent Sr, Ir and O atoms, respectively
  \cite{VESTA}. Red and white arrows on Ir atoms depict the directions of
  $J_{\rm eff}$ = 1/2 isospin moments, and blue arrows show the directions of
  net in-plane canting moments.}
\label{fig1}
\end{figure}

Despite the strong 2D character of Heisenberg exchange, Sr$_2$IrO$_4$ orders antiferromagnetically likely due to a small but finite interlayer coupling, which is also the case of La$_2$CuO$_4$. The magnetic structure of Sr$_2$IrO$_4$ was revealed by resonant x-ray magnetic scattering  \cite{BJ_Science2009} as illustrated in Fig.~\ref{fig1}. Below $T_{\rm N}$, the $J_{\rm eff}$ = 1/2 isospins lying in the basal planes form a N${\rm \acute{e}}$el order. Since the crystal structure of Sr$_2$IrO$_4$ has the staggered rotations of IrO$_6$ octahedra about the $c$-axis ($\sim11^{\circ}$) \cite{Huang_JSSC1994}, Dzyaloshinsky-Moriya (DM) interaction with ${\bf D}$ parallel to the $c$-axis is present \cite{Jackeli_PRL2009}, leading to canting of isospins and the appearance of small in-plane moment. The in-plane canting moments are cancelled out at zero field by forming the up-up-down-down (uudd) stacking configuration along the $c$-axis [Fig.~\ref{fig1}(a)], while at a field above $\mu_{0}H_{\rm c} \sim$ 0.2 T the in-plane moments align and produce weak-ferromagnetism with a moment $M$ of $\sim 0.075 \mu_{\rm B}$/Ir \cite{BJ_Science2009}. This magnetic structure of Sr$_2$IrO$_4$ resembles with that of La$_2$CuO$_4$. In La$_2$CuO$_4$, $S$ = 1/2 spins order antiferromagnetically, and the buckling distortion of CuO$_6$ produces in-plane canting moments normal to the CuO$_2$ planes through DM interaction, which stack antiferromagnetically along the $c$-axis. 

The uudd configuration of canting moments in Sr$_2$IrO$_4$ at a glance would suggest the presence of two different interlayer couplings between the neighboring IrO$_2$ planes. Considering the crystal structure, however, the interlayer couplings between the adjacent planes are all equivalent. In order to account for the uudd configuration, the interlayer couplings beyond the nearest neighbors must be taken into account. 

Thio et al. formulated the magnetic susceptibility $\chi (T)$ of La$_2$CuO$_4$ by a mean-field approximation with the DM interaction and the interlayer coupling \cite{Thio_PRB1988,Thio_PRL1994}, which well reproduced the experimental data, and confirmed the predominant 2D Heisenberg exchange in the CuO$_2$ planes. Sr$_2$IrO$_4$ inherits stronger DM interaction due to spin-orbit coupling of Ir, as evidenced by the much larger canting moment compared with that of La$_2$CuO$_4$ ($M$ $\sim 2 \times 10^{-3} \mu_{\rm B}$/Cu) \cite{Thio_PRB1988}. As a critical test for the 2D IHAF, similar mean-field analysis on Sr$_2$IrO$_4$ is desired.

In this paper, we present the analysis of  $\chi (T)$ of Sr$_2$IrO$_4$ in the paramagnetic phase above $T_{\rm N}$. We formulated the theoretical magnetic susceptibility based on the Heisenberg model incorporating DM interaction and introduced the interlayer couplings within a mean-field approximation. The experimental data were fitted by the theoretical susceptibility, and the analysis indicates that  $\chi (T)$ is fully consistent with the predominance of isotropic Heisenberg exchange. The parameters obtained by the fit point to a competing nature of nearest and next-nearest interlayer couplings.

\section{Experimental}

Single crystals of Sr$_2$IrO$_4$ were grown by using SrCl$_2$ flux \cite{BJ_Science2009}. Magnetization data were collected by a commercial magnetometer (Quantum Design, MPMS). In order to obtain sizable magnetization signal at low fields, $\sim$20 thin plate-like single crystals ($\sim$2 mm $\times$ 2 mm $\times$ 0.05 mm) were piled up to form a block-shaped sample ($\sim$18 mg). The background contribution from the sample holder was measured independently, and was subtracted from the raw magnetization data.

\section{Results}

\begin{figure}
\epsfysize=75mm
\centerline{\epsffile{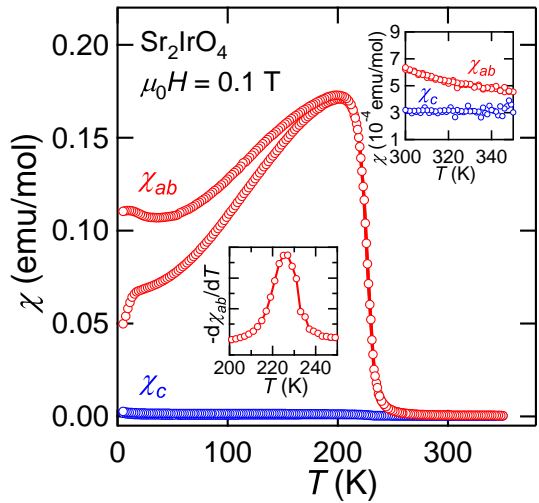}}
\caption{Temperature dependence of in-plane ($\chi_{ab}$) and out-of-plane ($\chi_c$) magnetic susceptibilities of Sr$_2$IrO$_4$ measured at 0.1 T. The lower inset shows the temperature derivative of in-plane susceptibility which shows a peak at $\sim$230 K, and the upper one displays the magnified view of high-temperature region.}
\label{fig2}
\end{figure}

The temperature dependent magnetic susceptibility, measured at a low field of 0.1 T, is shown in Fig.~\ref{fig2}. A large anisotropy between the in-plane ($\chi_{ab} \; \equiv \; M_{ab}/H$) and the out-of-plane ($\chi_{c} \; \equiv \; M_{c}/H$) susceptibilities is clearly seen \cite{chi}. Only $\chi_{ab}$ displays a pronounced temperature dependence roughly below room temperature, while $\chi_{c}$ remains almost constant over the whole temperature range measured. The observed anisotropy should be attributed to the in-plane canting moments produced by DM interaction with ${\bf D}$ // the $c$-axis. 

In the in-plane susceptibility $\chi_{ab}$, a peak is observed around 200 K which is lower than $T_{\rm N}$ $\sim$ 230 K determined by a magnetic x-ray diffraction measurement \cite{Fujiyama_diffuse}. $T_{\rm N}$ appears to be reflected as the peak temperature in the temperature derivative of susceptibility, namely the temperature with the steepest slope in rapidly increasing susceptibility on cooling to the peak at 200 K (see the lower inset of Fig.~\ref{fig2}).  In contrast to the previous data measured at a relatively high field of 0.5 T \cite{Cao_PRB1998}, $\chi_{ab}$ shows a clear decrease with cooling below 200 K. This is consistent with the uudd stacking of canting moments in the ground state \cite{BJ_Science2009}, where the net moments are zero. The bifurcation seen well below $T_{\rm N}$ likely represents the uncompensated canting moments due to the pinning to crystalline defects such as stacking faults along the $c$-axis. We believe that the competition of very weak nearest and next-nearest interlayer couplings, as will be discussed below, is one of the origins for such pronounced pinning effect. 

\begin{figure}
\epsfysize=60mm
\centerline{\epsffile{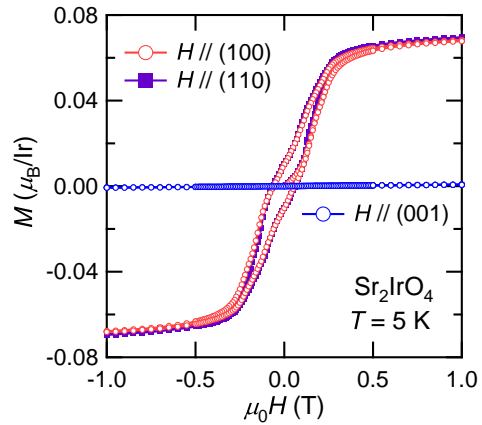}}
\caption{Isothermal magnetization curves of Sr$_2$IrO$_4$ registered at 5 K.}
\label{fig3}
\end{figure}

In the isothermal magnetization curve at 5 K well below $T_{\rm N}$, shown in 
Fig.~\ref{fig3}, a metamagnetic transition from an antiferromagnetic ground
state to a weak ferromagnetic state  can be seen \cite{Cao_PRB1998}. A sudden
increase of the in-plane magnetization at around $\mu_{0}H_{\rm c} \sim$ 0.15 T
was observed, which corresponds to the flipping of net in-plane moments as
illustrated in Fig.~\ref{fig1}. The reduced slope at the zero field limit mirrors the suppressed in-plane susceptibility below 200 K in the temperature dependent susceptibility. Measuring the magnetization above $\mu_{0}H_{\rm c}$ gives rise to a weakly ferromagnetic behavior as reported previously \cite{Cao_PRB1998}. The magnitude of weak ferromagnetic moments is $\sim$ 0.068 $\mu_{\rm B}$/Ir, slightly smaller than a reported value of $\sim$ 0.075 $\mu_{\rm B}$/Ir \cite{BJ_Science2009}. (100) orientation of moment (in the $\sqrt{\mathstrut 2}a \times \sqrt{\mathstrut 2}a$ unit cell where $a$ is the nearest Ir-Ir distance) is known to be realized in the ordered state under zero filed \cite{Dhital_PRB2013}. However, any appreciable anisotropy in the magnetization curve was not detected between the (100) and (110) directions as shown in Fig.~\ref{fig3}. The in-plane anisotropy should be finite but extremely small. We do not observe any trace of metamagnetism along the $c$-axis, consistent with the canting moments only within the $ab$-planes by ${\bf D}$ // the $c$-axis \cite{Mc_discussion}. Since the metamagnetism is associated with a change in the magnetic interlayer sequence along the $c$-axis, we can estimate the effective interlayer coupling energy as the product of the metamagnetic moment  $\Delta M_{ab}$ and the critical
magnetic field $H_{\rm c}$ as $\Delta M_{ab}\cdot \mu_{0} H_{\rm c} \sim 0.06 \;\mu_{\rm B} \times 0.15 \;{\rm T} \sim 0.7\; \mu$eV. There is a hysteresis in the magnetization at a low field region, which shows up as the bifurcation in the temperature dependent susceptibility and should be extrinsic. 

With increasing temperature above $T_{\rm N}$, $\chi_{ab}$ decreases and appears to crossover to almost temperature independent behavior. The magnitude of $\chi_{ab}$ in the high temperature limit is comparable to $\chi_{c}$ as seen in the upper inset of Fig.~\ref{fig2}, implying that the isospin system is isotropic in the paramagnetic phase. We will analyze this region in detail as a weakly coupled 2D Heisenberg system with DM interaction. 
 
\section{Discussion}
\subsection{Fitting of $\chi(T)$ based on 2D IHAF model.}

The interlayer coupling energy is orders of magnitude smaller than that of the in-plane coupling characterized by the in-plane antiferromagnetic coupling $J_{ab}$ $\sim$ 0.1 eV \cite{Fujiyama_diffuse,Lemmens_Sr2IrO4}. This should give rise to strong 2D magnetic fluctuations over a wide temperature range up to $\sim J_{ab}/k_{\rm B}$ well above the three-dimensional ordering temperature $T_{\rm N}$.  The Heisenberg character of 2D fluctuations was captured as the temperature dependence of magnetic correlation length above $T_{\rm N}$ measured by a resonant x-ray diffuse scattering \cite{Fujiyama_diffuse}.  As described in the introduction, the magnetism of $J_{\rm eff}$ = 1/2 isospins is in striking parallel with the case for 2D $S$ = 1/2 Heisenberg antiferromagnet La$_2$CuO$_4$. The two-dimensional magnetic correlations ($J_{ab}$ $\sim$ 0.135 eV \cite{Thio_PRL1994}) first develop on cooling, and the finite interlayer coupling ($\sim 1\; \mu$eV) triggers the three-dimensional magnetic ordering at $T_{\rm N}$ \cite{Keimer_PRB1992,Thio_PRB1988}. 

In accord with the close analogy of $J_{\rm eff}$ = 1/2 magnetism of Sr$_2$IrO$_4$ with 2D $S$ = 1/2 IHAF in La$_2$CuO$_4$, we emphasize here that the temperature dependent magnetic susceptibility $\chi (T)$ of La$_2$CuO$_4$ is surprisingly similar to that of Sr$_2$IrO$_4$. The out-of-plane susceptibility $\chi_{c}$ of La$_2$CuO$_4$ displays a sharp peak at $T_{\rm N}$ while the in-plane susceptibility $\chi_{ab}(T)$ shows only a very weak temperature dependence \cite{Thio_PRB1988, La2CuO4_Cheong}. A clear signature of metamagnetism was observed below $T_{\rm N}$ in the out-of-plane magnetization curve, evidencing the presence of canting moments \cite{Thio_PRB1988,Thio_PRL1994,La2CuO4_Cheong}.

In La$_2$CuO$_4$, the steep increase of $\chi_{c}(T)$ right above $T_{\rm N}$ is attributed to the canting moment produced by DM interaction in the presence of developed 2D magnetic correlations. The theoretical magnetic susceptibility formulated by Thio et al., which is based on 2D IHAF incorporating DM interaction and interplayer coupling, well described  $\chi (T)$ of La$_2$CuO$_4$  \cite{Thio_PRB1988,Thio_PRL1994}.  In the following, we attempt to describe $\chi (T)$ of Sr$_2$IrO$_4$ in the same framework.

In order to formulate $\chi(T)$ of Sr$_2$IrO$_4$, two major differences from La$_2$CuO$_4$ must be taken into account. (i) In Sr$_2$IrO$_4$, the rotation of IrO$_6$ octahedra about the $c$-axis gives rise to the DM vector parallel to the $c$-axis, whereas the buckling of CuO$_6$ along (010) (in the $\sqrt{2}a \times \sqrt{2}a$  unit cell) yields the DM vector lying in the CuO$_2$ plane. This results in the direction of the canting moments parallel to the IrO$_2$ plane in Sr$_2$IrO$_4$, while that is perpendicular to the CuO$_2$ plane in La$_2$CuO$_4$. (ii) The interlayer coupling is dominated by nearest neighbor antiferromagnetic interaction in La$_2$CuO$_4$. For Sr$_2$IrO$_4$, the interlayer couplings beyond the nearest neighboring planes must be considered to allow for the uudd configuration of canting moments. With these differences in mind, we construct the theoretical magnetic susceptibility of Sr$_2$IrO$_4$ in the paramagnetic phase.

\begin{figure}
\epsfysize=70mm
\centerline{\epsffile{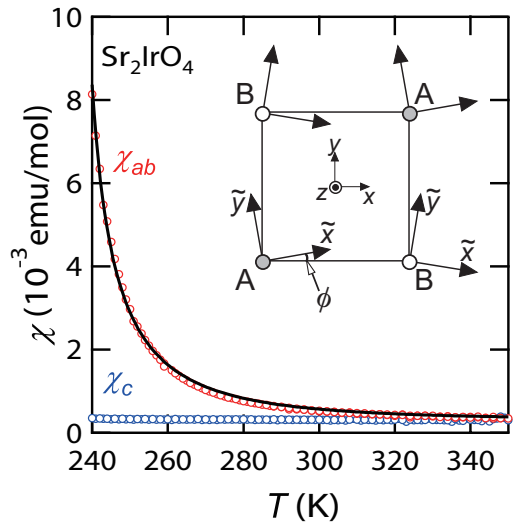}}
\caption{Magnetic susceptibilities of Sr$_2$IrO$_4$ in the high-temperature
  paramagnetic phase. Red and blue dots show the experimental in-plane
  ($\chi_{ab}$) and out-of-plane ($\chi_{c}$) susceptibilities, and the black
  solid line delineates the fitting line for $\chi_{ab}$ based on Eq. (4). The
  inset depicts the local spin frame ($\tilde{x}$, $\tilde{y}$) rotated by the
  angle of $\pm\phi$ from the laboratory frame of ($x$, $y$). A and B
  represent the two antiferromagnetic sublattices.}
\label{fig4}
\end{figure}

To derive the theoretical magnetic susceptibility, we introduce the local axes for A and B magnetic sublattices. They are obtained by a staggered rotation of spin-axis about $z$-axis (i.e. the crystallographic $c$-axis) with angles of $\pm\phi$, as sketched in the inset of Fig.~\ref{fig4}. In the rotated axis frame, the intralayer magnetic coupling can be mapped onto IHAF, if we ignore the Hund's coupling, as discussed in Ref.~\cite{Jackeli_PRL2009}. By introducing the interlayer couplings in a mean-field approximation, the in-plane susceptibility of 3D coupled layers is expressed as follows in terms of its out-of-plane component $\chi_{c}$ and staggered susceptibility of 2D IHAF $\chi^{\dagger}$ given in units of inverse energy. We find (see Appendix),

\begin{eqnarray}
\chi_{ab}=
\cos^2\phi\chi_{c}+\frac{\sin^2\phi(g_{ab}\mu_{B})^2{\chi}^{\dagger}}
{1-J_c{\chi}^{\dagger}}~
\label{eq1}
\end{eqnarray}
where $\mu_{\rm B}$ and $g_{ab}$ denote respectively Bohr magnetron and the in-plane $g$-factor of $J_{\rm eff}$ = 1/2 isospin, which is 2 in the cubic limit \cite{BJ_PRL2008}. For the interlayer couplings, we first consider a single parameter $J_c$ which represents an effective exchange field coming from all interlayer exchange couplings.

We analyze the experimental in-plane susceptibility in the paramagnetic phase shown in Fig.~\ref{fig4} based on Eq. (\ref{eq1}). The out-of-plane susceptibility $\chi_{c}$ is independent of temperature in the range shown in Fig.~\ref{fig4}, and estimated to be 3.1 $\times$ 10$^{-4}$ emu/mol \cite{J_NN_discussion}. Makivic and Ding studied the $S$ = 1/2 2D Heisenberg model on a square lattice by quantum Monte Carlo simulation, and obtained the following relation for staggered susceptibility $\chi^{\dagger}$,

\begin{eqnarray}
\chi^{\dagger} = 1.65(\xi/a)^{2}(k_{\rm B}T/J_{ab}^2)
\label{eq2}
\end{eqnarray}
where $\xi$ is the two-dimensional magnetic correlation length and $a$ is the nearest Ir-Ir distance \cite{MD_MC}. $\xi$ is expressed as, 

\begin{eqnarray}
\xi = 0.276a {\rm exp}(1.25J_{ab}/T)
\label{eq3}
\end{eqnarray}
which well explained the experimental data obtained by resonant x-ray diffuse scattering and yielded $J_{ab}$ as 0.1 $\pm$ 0.01 eV \cite{Fujiyama_diffuse}. We note that $\chi^{\dagger}$ estimated from Eq. (\ref{eq2}) agrees well with the one obtained by a large scale quantum Monte Carlo method \cite{Kim_Troyer}.

Throughout the analysis, we employed the following assumptions so as to obtain a reliable fit. (i) Since $\phi$ is at most $\sim11^\circ$, which is the angle of IrO$_6$ rotations about the $c$-axis, ${\rm cos}^{2}\phi$ should be $0.97 < {\rm cos}^2\phi < 1$, namely very close to 1. We thus omitted the prefactor of ${\rm cos}^{2}\phi$ for the out-of-plane susceptibility $\chi_{c}$. We confirmed that the presence or absence of this factor did not alter the final results \cite{Supplemental}. (ii) Since $g_{ab}$ and $\phi$ cannot be determined independently, we treated $g_{ab}{\rm sin}\phi$ as a single parameter. By taking $J_c$ and $g_{ab}{\rm sin}\phi$ as variant parameters and fixing $J_{ab}$ at 0.10 eV, we fitted the experimental in-plane susceptibility in the temperature range between 240 K and 350 K \cite{Supplemental}.

The result of fit is shown as the black solid line in Fig.~\ref{fig4}. Eq. (\ref{eq1}) reasonably reproduces the experimental result, indicating the predominance of IHAF in Sr$_2$IrO$_4$. This also implies that the influence of pseudodipolar interaction induced by Hund's coupling is not appreciable in the high temperature paramagnetic phase of Sr$_2$IrO$_4$ \cite{Hund}. The obtained parameters are $g_{ab}{\rm sin}\phi = 0.0376 \pm 0.0002$ and $J_{c} = 15.86 \pm 0.07 \mu$eV. We note that a steep increase of $\chi_{ab}$ right above $T_{\rm N}$ by DM interaction benefits in obtaining a reliable fit. This contrasts with a sister compound Ba$_2$IrO$_4$ where DM interaction is absent and its magnetic susceptibility shows no visible anomaly at $T_{\rm N}$ \cite{Isobe_Ba2IrO4}.

The obtained $g_{ab}{\rm sin}\phi \sim 0.038$ is small compared with the one estimated from the weak ferromagnetic moment in the ordered state at low temperatures, $M = g_{ab}\mu_{\rm B}S{\rm sin}\phi \sim 0.068 \;\mu_{B}$. For $S$ = 1/2, the $g_{ab}{\rm sin}\phi \sim 0.13$, a factor of 3 larger than the fitting result. Since the isospin moments were found to rigidly follow the IrO$_6$ rotations \cite{Bosseggia_JPhys2013} and the change of IrO$_6$ rotation angle is less than 1$^{\circ}$ between room temperature and 10 K \cite{Crawford_PRB49}, we cannot ascribe the difference to the change of isospin canting angle by temperature. The possible origin of this discrepancy is the reduced magnitude of isospin moments at high temperatures. Due to the smallness of charge gap of $\sim$ 0.5 eV \cite{BJ_PRL2008}, charge excitation is substantial at high temperatures which may renormalize the size of effective local moment. Such renormalization might be a characteristic feature of weak Mott insulators with a small charge gap.

\subsection{Up-up-down-down stacking configuration of in-plane canting moments.}

The fitting result shows an effective antiferromagnetic interlayer coupling $J_{c} > 0$. If only the nearest plane interlayer coupling is considered, this cannot lead to the uudd interlayer sequence of in-plane moments, and the interlayer couplings beyond nearest planes must be taken into account. The presence of sizable further neighbor interlayer couplings should be reasonable in the sence that Sr$_2$IrO$_4$ is regarded as a weak Mott insulator marginally formed by modest Coulomb $U$ of 5$d$ electrons. Since the interlayer couplings beyond the next-nearest neighboring planes are supposed to be negligibly small, we consider the interlayer couplings from the nearest and next-nearest planes. 

\begin{figure}
\epsfysize=60mm
\centerline{\epsffile{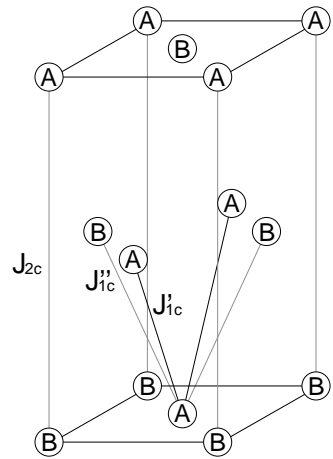}}
\caption{Interlayer couplings in nearest
  ($J^{\prime}_{1c}$ and $J^{\prime\prime}_{1c}$) and next-nearest
  ($J_{2c}$) planes.
The iridium sublattice formed by anticlockwise (clockwise) rotated
  octahedra is labeled by $\mathsf{A}$ ($\mathsf{B}$).}
\label{fig5}
\end{figure}

We construct a minimal model that includes the isotropic couplings between iridium ions in nearest
($J^{\prime}_{1c}$ and $J^{\prime\prime}_{1c}$ within the same and different sublattices, respectively) and next-nearest ($J_{2c}$) planes (see Fig.~\ref{fig5}).  The fact that there is no visible anisotropy in the measured in-plane magnetizations justifies to drop out symmetry allowed anisotropy terms. By introducing the planar unit vector ${\vec m}_n$ for the staggered moment of $n$th-layer, which corresponds to $({\vec S}_{{\rm A}, n} - {\vec S}_{{\rm B}, n})/2S$ where ${\vec S}_{{\rm A(B)}, n}$ denotes spins of A(B) sublattice of $n$th-layer, we arrive to the following classical energy (in unit of $1/S^{2}$) of coupled layers,

\begin{align}
E=\sum_{n} \{-j_{1c}{\vec m}_n\cdot{\vec m}_{n+1}+j_{2c}{\vec
  m}_n\cdot{\vec m}_{n+2}\nonumber\\
-b({\vec m}_n\cdot{\vec m}_{n+1})^2\}~,
\label{eq4}
\end{align}
where $j_{1c}=2(J_{1c}^{\prime\prime}\cos2\phi-J_{1c}^{\prime})$ and
$j_{2c}=-J_{2c}\cos2\phi$ are effective exchange couplings between nearest and
next-nearest neighbor planes, respectively \cite{sign_interlayer}. 
In addition to the classical energy, we also include effective biquadratic
coupling $b$ driven by  quantum fluctuations \cite{Yildirim_PRL1994}.
In order to allow for the uudd configuration, we assume ferromagnetic (antiferromagnetic) nearest (next-nearest) interlayer couplings between ${\vec m}_n$, respectively \cite{sign_interlayer}.

The effective interlayer coupling $J_c$, entering in Eq. (\ref{eq1}), is expressed as $J_{c} = 2(j_{1c}-j_{2c}$). From the fit value of $J_{c} \sim 15.9\; \mu$eV and the change of exchange energy at the the field induced metamagnetic transition $(2j_{1c} - j_{2c})S^{2} \; = \; \Delta M_{ab}\cdot \mu_{0} H_{\rm c}$ $\sim 0.7\; \mu$eV, we obtain $j_{1c}$ and $j_{2c}$ as 18.8 $\mu$eV and 10.8 $\mu$eV, respectively. The result points to a competition of nearest and next-nearest interlayer couplings with a comparable magnitude, which originates from the following facts; (i) $j_{1c}$ is geometrically frustrated, that is, $J^{\prime}_{1c}$ and $J^{\prime\prime}_{1c}$ compete each other and $j_{1c}$ is marginally yielded by the uncompensated out-of-plane exchanges due to the rotational distortion of IrO$_6$ ochtahedra. (ii) The cubic character of $J_{\rm eff}$ = 1/2 wave function, in contrast to the outermost Cu $d_{x^{2}-y^{2}}$ orbital of La$_2$CuO$_4$ extended in the basal plane, gives rise to sizable $J_{2c}$.

The result accounts for the uudd configuration of in-plane moments observed by the diffraction experiments \cite{BJ_Science2009,Dhital_PRB2013}. The uudd configuration is stabilized in the following parameter range of the present model [Eq. (\ref{eq4})] \cite{Kaplan_PRB2009},

\begin{eqnarray}
2j_{2c} > j_{1c} > 0 \;\;{\rm and}\;\; b > \frac{j_{1c}^2}{4(2j_{2c} - b)} > 0.
\label{eq5}
\end{eqnarray}

$j_{1c}$ and $j_{2c}$ obtained as above satisfy the former relation. The latter relation calls for $b > 5.5 \;\mu$eV, which is three orders of magnitude larger than the value reported for La$_2$CuO$_4$ ($b \sim 2 \times 10^{-9}$ eV) \cite{Yildirim_PRL1994}. Since $b$ is in proportion to $J_{\rm out}^{2}S/J_{ab}$, where $J_{\rm out}$ is the isotropic exchange between nearest neighbors in the adjacent planes ($J_{\rm out} \sim J^{\prime}_{1c}$ or $J^{\prime\prime}_{1c}$), this suggests the larger interlayer coupling $J_{\rm out}$ and thus larger interlayer hopping $t_{\bot}$ in Sr$_2$IrO$_4$. In a crude estimation \cite{estimate}, we obtain $t_{\bot}$(Sr$_2$IrO$_4$) $\sim$ 5 $t_{\bot}$(La$_2$CuO$_4$). The larger $t_{\bot}$ of Sr$_2$IrO$_4$ is consistent with the smaller anisotropy of resistivity $\rho_{c}/\rho_{a}$ of Sr$_2$IrO$_4$ than that of La$_2$CuO$_4$ \cite{FWang_PRL2011}, again attributed to the weak Mott character and the cubic shape of $J_{\rm eff}$ = 1/2 wave fucntion. The uudd configuration of in-plane moments in Sr$_2$IrO$_4$ is therefore stabilized by the following factors; (i) geometrically frustrated nature of nearest neighbour interlayer couplings which suppresses $j_{1c}$, (ii) isotropic and extended character of $J_{\rm eff}$ = 1/2 wave function giving rise to sizable $j_{2c}$ and $b$.
 
 \section{Conclusion}

We analyzed the magnetic susceptibility of spin-orbital Mott insulator Sr$_2$IrO$_4$ in the paramagnetic phase. The analysis evidences the predominance of isotropic Heisenberg exchange between the $J_{\rm eff}$ = 1/2 isospins, further reinforceing the similarity with La$_2$CuO$_4$. The result of fit points to the competing interlayer couplings between the nearest and next-nearest IrO$_2$ planes. The competing nature of interlayer couplings and the resultant complex stacking pattern of $J_{\rm eff}$ = 1/2 isospin moments might give a clue for further unsettled issues of Sr$_2$IrO$_4$ such as high-pressure suppression of weak-ferromagnetic moments \cite{Haskel_PRL} and the second magnetic transition below $T_{\rm N}$ argued from the local probes \cite{Sr2IrO4_muSR}.

\section*{Acknowledgements}

We are grateful to G. Khaliullin, V. Kataev, P. Lemmens and B. Keimer for invaluable discussion and R. Kremer for technical support. This work is partly supported by Grant-in-Aid for Scientific Research (S) (Grant No.24224010) and Grant-in-Aid for Scientific Research on Innovative Areas (Grant No. JP15H05852) from JSPS of Japan. G.J. is supported in part by the National Science Foundation under Grant No. NSF PHY11-25915.

\section*{Appendix}
In this Appendix, we derive the in-plane susceptibility of 3D coupled layer system in terms of uniform and staggered susceptibilities of 2D IHAF. The exchange interactions for an intralayer bond of
nearest-neighbor  iridium ions can be written as: 
\begin{align}\tag{A1}
{\cal H}_{ij}=J\vec S_i\cdot \vec S_{j}+\Gamma S_{i}^{z}S_{j}^{z}
-D{\big (}S_i^xS_j^y-S_i^yS_j^x{\big )}~,
\label{eqS1}
\end{align}
they include isotropic antiferromagnetic (AF) coupling ($J$), 
as well as symmetric ($\Gamma$) and antisymmetric ($D$) Dzyaloshinsky-Moriya (DM) exchange anisotropies  
\cite{Jackeli_PRL2009}. The dominant contributions (ignoring Hund's exchange induced
corrections) to the coupling constants in Eq.~\ref{eqS1} can be 
parameterized as $J={\tilde J}\cos 2\phi$, $\Gamma=2{\tilde J}\sin^2\phi$,
and $D={\tilde J}\sin 2\phi$, where ${\tilde J}=\sqrt{J^2+D^2}$ 
defines overall energy scale and $\tan2\phi=D/J$.
Following Ref.~\cite{Jackeli_PRL2009}, we introduce the local quantization axes for
spins on $\mathsf{A}$ and $\mathsf{B}$ sublattices obtained by a staggered rotation of
the spin frame around $z$-axis by an angle $\pm\phi$ 
[see inset in Fig.~\ref{fig4} of
the main text]. We further denote by ${\tilde S}_{i}^{\gamma}$ ($\gamma=x,~y,~z$)
the Cartesian components of spins in a local rotated frame. They are
related to the laboratory frame by the following transformations:
\begin{align}
&S_{i}^{x}=\cos\phi{\tilde S}_{i}^{x}-\exp(\imath{\bf Q}{\bf R}_{i})\sin\phi{\tilde S}_{i}^{y}~,\tag{A2}\label{eqS2}\\
&S_{i}^{y}=\cos\phi{\tilde S}_{i}^{y}+\exp(\imath{\bf Q}{\bf R}_{i})\sin\phi{\tilde S}_{i}^{x}~,~
S_{i}^{z}= {\tilde S}_{i}^{z}~.\nonumber
\end{align}
Here ${\bf Q}=(\pi,\pi)$ and $\exp(\imath{\bf Q}{\bf R}_{i})=+(-)1$
for $i$ belonging to $\mathsf{A}$ ($\mathsf{B}$) sublattice. With this transformation, the
anisotropic Hamiltonian Eq.~\ref{eqS1}, with the above parameterization of coupling
constants, is mapped in the rotated frame to  the isotropic
Heisenberg AF (IHAF) 
\begin{align}
{\tilde {\cal H}}_{ij} =
{\tilde J}\vec {\tilde S}_i\cdot\vec{\tilde S}_{j}.
\tag{A3}
\label{eqS3}
\end{align}

Thus in the rotated frame spins form collinear N\'eel order.
The Hund's coupling induced anisotropy selects in-plane AF order 
(see Ref.~\cite{Jackeli_PRL2009}), and corresponding spin pattern in the laboratory frame 
is given by canted AF structure with canting angle $\phi$ 
[see inset in Fig.~\ref{fig4} of the main text].

Based on the above derived mapping, we  relate the magnetic
susceptibilities of the system described by anisotropic Hamiltonian
Eq.~\ref{eqS1} to that of isotropic IHAF Eq.~\ref{eqS3}. To this end, we 
first rewrite the transformation Eq.~\ref{eqS2} in the momentum
representation
\begin{align}
&S_{{\bf q}}^{x}=\cos\phi{\tilde S}_{\bf q}^{x}-\sin\phi{\tilde S}_{{\bf q}+{\bf Q}}^{y}~,~\tag{A4}\label{eqS4}\\
&S_{{\bf q}}^{y}=\cos\phi{\tilde S}_{\bf q}^{y}+\sin\phi{\tilde S}_{{\bf q}+{\bf Q}}^{x}~,~
S_{\bf q}^{z}= {\tilde S}_{\bf q}^{z}~,\nonumber
\end{align}
we then express the in-plane and out-of-plane (along the
$c$-axis) components of uniform static magnetic susceptibility of a single plane, $\chi_{ab}$ and $\chi_{c}$ respectively, modeled by
Eq.~\ref{eqS1} in terms of uniform 
$\chi_{0}=\chi({\bf q}=0)$ and staggered 
$\chi^{\dagger}=\chi({\bf q}={\bf Q})$
static susceptibilities of 2D IHAF:
\begin{align}
\chi_{ab}=
\cos^2\phi\chi_{0}+\sin^2\phi\chi^{\dagger}~,~
\chi_{c}=\chi_{0}~.\tag{A5}
\label{eqS5}
\end{align}

It is straightforward to generalize  Eq.~\ref{eqS5}  to a 3D  system of coupled layers,
 such as  Sr$_2$IrO$_4$ of interest here,   we find
\begin{align}
\chi_{ab}=
\cos^2\phi\chi_{0}+\sin^2\phi\chi^{\dagger}_{+}~,~
\chi_{c}=\chi_{0}~,\tag{A6}
\label{eqS8}
\end{align}
where $\chi^{\dagger}_{+}$ now stands for the susceptibility of coupled layers in response to the applied field modulated in such a way that each $\mathsf{A}$
sublattice of different layers influence the same field that is opposite to the
one influenced by $\mathsf{B}$ sublattices. 
We next relate  $\chi^{\dagger}_{+}$ to  $\chi^{\dagger}$ (staggered
susceptibility of 2D IHAF) within the random phase approximation (RPA) 
[see e.g. Ref.~\cite{Sca75S}] for the weak interlayer couplings. We find

\begin{align}
&{\chi}^{\dagger}_{+}=\frac{{\chi}^{\dagger}}{1-J_{c}{
    \chi}^{\dagger}}~,\tag{A7}\label{eqS9}\\
&\chi_{ab}=
\cos^2\phi\chi_{c}+\frac{\sin^2\phi(g_{ab}\mu_{B})^2{\chi}^{\dagger}}
{1-J_c{\chi}^{\dagger}}~\nonumber
\end{align}
where $J_c$ is an effective exchange field  from neighboring
layers, and $\chi^{\dagger}_{+}$ and $\chi^{\dagger}$ are given in units of inverse energy. Note that biquadratic coupling $b$ does not contribute to a linear
susceptibility of paramagnetic state.



\begin{thebibliography}{99}
\bibitem{chi}
Although the magnetization is non-linear against magnetic field below $T_{\rm N}$, we define $\chi$ as $M/H$ throughout this paper.
\bibitem{Mc_discussion}
Only very tiny fraction of ferromagnetic moments (a few percent of $M_{ab}$) was seen along the $c$-axis, highly likely due to small misalignment of the sample in the magnetization measurement. We could reasonably subtract the ferromagnetic admixture from $M_c$ by evaluating the in-plane contribution as $M_{ab}H{\rm sin}\theta\;\times\;{\rm sin}\theta$, where $\theta$ denotes the misalignment angle estimated to be $\sim3^{\circ}$.
\bibitem{J_NN_discussion}
By subracting the contributions of core diamagnetism, the uniform susceptibility $\chi_{0}$ is estimated to be 4.2 $\times$ $10^{-4}$ emu/mol. From the relation $\chi_0 = (g\mu_{B})^2/2zJ_{ab}$ where $z$ is the number of nearest neighbors and $z = 4$ for Sr$_2$IrO$_4$, we obtain $J_{ab}$ as $\sim$ 40 meV, much smaller than the value estimated from the temperature dependence of magnetic correlation length \cite{Fujiyama_diffuse}. We presume that this discrepancy is attributed to the contribution of van Vleck susceptibility to the experimental $\chi_{c}$, which is of the order of $10^{-4}$ emu/mol in iridates \cite{CJK_zigzag}.
\bibitem{Hund}
Even though the anisotropy by Hund's coupling is not appreciable at high temperatures, it serves to stabilize the in-plane canted antiferromagnetic structure rather than the collinear N\'eel ordering with isospins directed along the $c$-axis (The two magnetic structures are degenerate without Hund's coupling).
\bibitem{sign_interlayer}
We cannot determine the signs of $J^{\prime}_{1c}$ and $J^{\prime\prime}_{1c}$ in this model, and the only required constraint is $j_{1c}$ $>$ 0 for the ferromagnetic coupling of ${\vec m}_n$ and ${\vec m}_{n+1}$. For the next-nearest interlayer coupling, $J_{2c}$ must be ferromagentic ($<$ 0) to realize the antirromagnetic configuration of ${\vec m}_n$ and ${\vec m}_{n+2}$ [see Fig.~\ref{fig1}(a)], and we thus put minus sign in $j_{2c} = -J_{2c}{\rm cos}2\phi$.
\bibitem{estimate}
In the estimate, we assumed $J_{ab}$ = 0.13 eV and Coulomb $U$ = 5 eV for La$_2$CuO$_4$ and $J_{ab}$ = 0.10 eV and $U$ = 2 eV for Sr$_2$IrO$_4$. $t_{\bot}$ is obtained from $J_{\rm out} = t_{\bot}^2 /U$, and $S$ is taken as 1/2 for both compounds.
\end{thebibliography}
\end{document}